\documentclass{article}
\usepackage{aimc2025}

\usepackage[utf8]{inputenc}
\usepackage[T1]{fontenc}
\usepackage{hyperref}
\hypersetup{
  colorlinks   = true,
  urlcolor     = blue,
  linkcolor    = black,
  citecolor   = black
}
\usepackage{url}
\usepackage{booktabs}
\usepackage{amsfonts}
\usepackage{nicefrac}
\usepackage{graphicx}
\usepackage{musicography}
\usepackage{amsmath, amssymb, amsthm}
\usepackage{multirow}
\usepackage{float}
\usepackage{enumitem}
\usepackage{tabularx}
\usepackage{bm}
\newcommand{\vectorize}[1]{\bm{#1}}
\usepackage{lineno}
\usepackage{microtype}
\usepackage{subcaption}
\usepackage{wrapfig}
\usepackage{array}
\newcolumntype{C}[1]{>{\centering\arraybackslash}p{#1}}
\newcommand{\val}[2]{#1 \pm \text{\scriptsize #2}}

\usepackage[table]{xcolor}

\newcommand{\others}[1]{\textcolor{lightgray}{#1}}
\definecolor{darkgreen}{HTML}{189c18}

\newcommand{\hlred}[1]{\textcolor{red}{#1}}
\newcommand{\hlblue}[1]{\textcolor{blue}{#1}}
\newcommand{\hlgreen}[1]{\textcolor{darkgreen}{#1}}
\newcommand{\hlorange}[1]{\textcolor{orange}{#1}}
\newcommand{\hlpurple}[1]{\textcolor{purple}{#1}}
\newcommand{\hlcyan}[1]{\textcolor{cyan}{#1}}
\newcommand{\hlmagenta}[1]{\textcolor{magenta}{#1}}

\title{From Generality to Mastery: Composer-Style Symbolic Music Generation via Large-Scale Pre-training}

\author{
  Mingyang Yao \\
  Department of Mathematics \\
  University of California, San Diego\\
  \texttt{m5yao@ucsd.edu} \\
  \And
  Ke Chen \\
  Department of Computer Science and Engineering \\
  University of California, San Diego\\
  \texttt{knutchen@ucsd.edu} \\
}

\begin{document}

\maketitle

\begin{abstract}
Despite progress in controllable symbolic music generation, data scarcity remains a challenge for certain control modalities. Composer-style music generation is a prime example, as only a few pieces per composer are available, limiting the modeling of both styles and fundamental music elements (e.g., melody, chord, rhythm). In this paper, we investigate how \textbf{general} music knowledge learned from a broad corpus can enhance the \textbf{mastery} of specific composer styles, with a focus on piano piece generation. Our approach follows a two-stage training paradigm. First, we pre-train a REMI-based music generation model on a large corpus of pop, folk, and classical music. Then, we fine-tune it on a small, human-verified dataset from four renowned composers, namely Bach, Mozart, Beethoven, and Chopin, using a lightweight adapter module to condition the model on style indicators. To evaluate the effectiveness of our approach, we conduct both objective and subjective evaluations on style accuracy and musicality. Experimental results demonstrate that our method outperforms ablations and baselines, achieving more precise composer-style modeling and better musical aesthetics. Additionally, we provide observations on how the model builds music concepts from the generality pre-training and refines its stylistic understanding through the mastery fine-tuning.
\end{abstract}

\section{Introduction}
Composer-style music generation aims to create musical pieces that adhere to the compositional techniques of a specific composer, which reflects distinct preferences in motif and texture development, chord progressions, and structural organization~\citep{harmony, mushistory}. The ability to model different composer styles in music generation can enrich music education by providing diverse learning materials and enabling quantitative analysis of stylistic features in musical works.

While symbolic music generation has made rapid progress with neural networks~\citep{musecoco, compembellish, cp, notagen, musictransformer, openai2019musenet}, controlling abstract concepts, such as composer style, remains a challenge. A major challenge is data scarcity. Due to copyright restrictions, data collection difficulties, and the natural lifespan limitation of composers, only a limited number of pieces per composer are available for model training, leading to suboptimal generation performance.
% Furthermore, a composer may reveal different specialties in a wide range of works and periods, making it hard to effectively learn the comprehensive style characteristics. For example, Chopin Waltzes (e.g., Op.64, Op.69), sonatas (e.g., Op.35, Op.58), and etudes (e.g., Op.10, Op.25) have distinct melodies, textures, and rhythmic patterns~\cite{waltzstyle, etudestyle}. Under data constraints, previous models might not be able to capture a typical style of a composer or overfit on a particular subset of works. 

However, the general symbolic music collections are abundant. Many public-domain symbolic music datasets either contain large-scale corpora~\citep{long2024pdmx}, or focus on specific genres such as pop~\citep{cp, pop909, bang-etal-2024-piast}, jazz~\citep{Pfleiderer:2017:BOOK}, classical~\cite{drengskapur2021midi, dcmlab2020dcmlcorpora, kernhumdrum, pianomidi}, or include specific labels~\citep{emopia}. These datasets contain rich music patterns in melody, harmony, rhythm, repetition, and variation, which are fundamental elements shared across all genres under the principles of music theory. Inspired by these resources, an under-explored question arises: \textbf{how} can a large corpus of unlabeled music be leveraged for composer-style modeling? Furthermore, \textbf{what} distinguishes a model trained on general music collections from one trained on the collection of specific composers? This question can essentially be denoted as ``from Generality to Mastery" in the symbolic music generation task.

In this paper, we propose a pipeline to address the above questions. Our contributions can be summarized as follows:
\begin{itemize}[leftmargin=15pt,itemsep=1pt,topsep=2pt]
    \item We propose a two-stage training paradigm for a transformer-based composer-style music generation model. The first stage pre-trains the model on a large music corpus of diverse genres, and the second stage fine-tunes the model on a small dataset of works from target composers. 
    \item To better represent the composer's style, we extend the original REMI representation~\citep{remi} to support all common time signatures and adjust its resolutions. Then we introduce adapter modules to embed composer conditions during the generation process.
    \item Experimental results show that our model outperforms previous baselines and ablations in composer-style modeling and musicality. Additionally, we analyze how the model develops music concepts through generality pre-training and refines its stylistic understanding through Mastery fine-tuning.  
\end{itemize}

We present generation demos and additional experiments in the demo website\footnote{\url{https://generality-Mastery.github.io/}}. The code implementation and datasets is open-sourced in \url{https://github.com/AndyWeasley2004/Generality-to-Mastery}

\section{Related Work}

\subsection{Symbolic Music Encoding}
For different symbolic music generation models, various music encoding methods have been proposed to representation symbolic music in the generation process, from MIDI-Message events~\citep{musictransformer}, REMI events~\citep{remi}, compound word~\citep{cp}, functional representation~\citep{emo}, to continuous latent embeddings such as MusicVAE~\citep{roberts2018musicvae}, EC2-VAE~\citep{yang2019deep}, SketchVAE~\citep{chen2020musicsketchnet}, and PianoTree VAE~\citep{wang2020pianotree}. Several works propose specific encoding methods for learning a style of one specific composer, primarily on Bach~\citep{deep_bach, musecoco}. But scaling these models to arbitrary styles without detailed annotations and architecture changes could be difficult. As the same-period work to our paper, NotaGen\citep{notagen} explores the large-scale training of ABC-notation-based music collection, and it is able to generate music pieces with some composer styles as one prefix indicator. 

\subsection{Symbolic Music Generation and Understanding}
Transformer architectures conduct promising results in the symbolic music generation task via their ability to capture long-range dependencies in sequences~\citep{transformer, musictransformer}. MuseNet~\citep{openai2019musenet} was trained on a large-scale MIDI dataset covering various genres, instruments, and composers, achieving genre-blending generations and rudimentary style control via prompt tokens. Besides the generation tasks, some transformer-based symbolic music representation models, such as  MIDI-Bert, MusicBert, and MuseBert~\citep{musicbert, wang2021musebert, midibert}, adopt the BERT-like representation and training paradigms on the symbolic music data for the music understanding or music generation purposes. 

Unlike previous approaches that solely train the models on large-scale datasets for general music generation or understanding tasks, this paper combines pre-training on a diverse music corpus and fine-tuning on small, composer-specific datasets to investigate their combined impact, with particular focus on their understanding on the composer-style music generation.

\begin{figure}
    \centering
    \includegraphics[width=\textwidth]{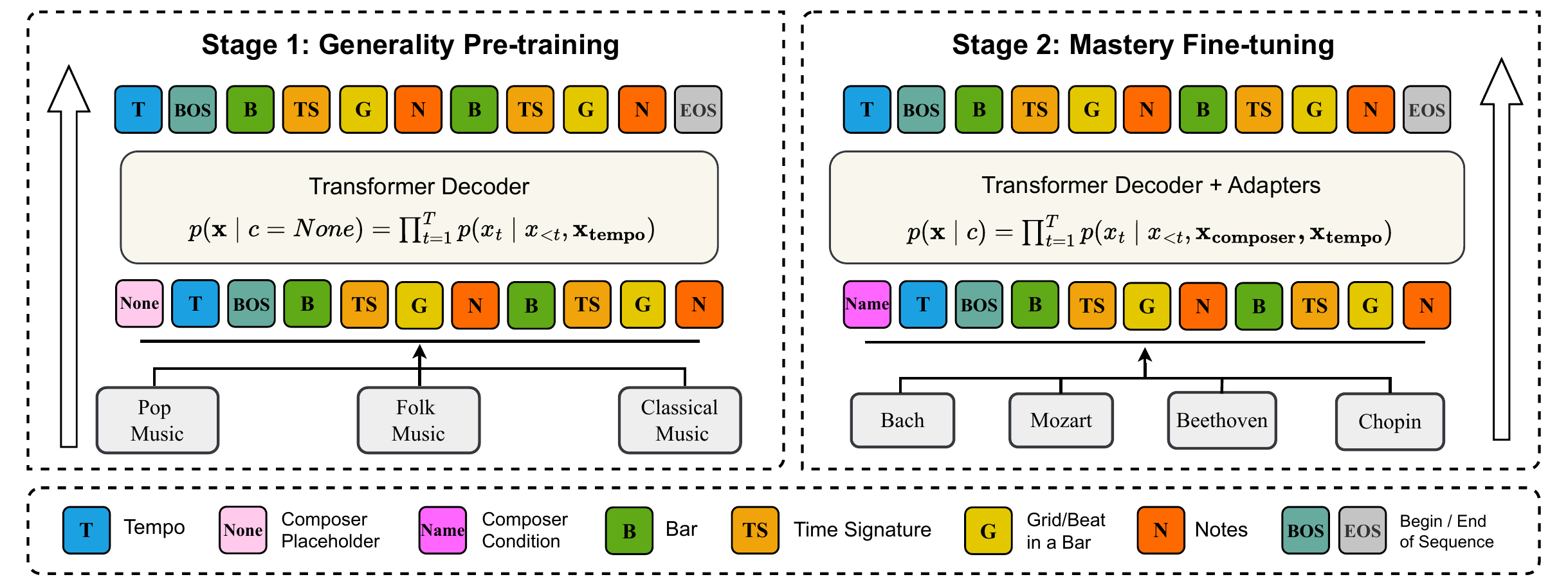}
    \caption{The pipeline of the proposed model \textit{GnM} (Generality and Mastery) for data-efficient composer-style symbolic music generation.}
    \label{fig:architecture}
\end{figure}

\section{Method}
As illustrated in Figure~\ref{fig:architecture}, our proposed model \textit{GnM} for composer-style music generation contains two training stages of the next-token prediction task on a transformer decoder model. The input of the model is a music sequence of extended REMI tokens, including: (1) global conditions, (2) structural cues, and most importantly (3) musical content. 
% After pre-training on generality and fine-tuning for Mastery, the model can produce sequences that are both structurally coherent and stylistically rich.

\subsection{Extended REMI Representation} 
Our representation stems from the original REMI~\citep{remi} with a few key extensions.

\subsubsection{Global Conditions}
We prepend each input sequence with a composer token \textbf{[Name]} and a tempo token \textbf{[T]} in the beginning. To avoid the large variations of the song tempos, we quantized the tempo into four categories (40, 80, 120, 160 BPM) to provide a simplified global speed condition. \textbf{[BOS]} and \textbf{[EOS]} denote the beginning and ending points, respectively. Notably, since each song will be divided into different segments as input sequences, we only assign \textbf{[BOS]} on the \textbf{true beginning point} of the song, and \textbf{[EOS]} on the \textbf{true ending point} of the song. 

\subsubsection{Structural Conditions}
Similar to~\citep{figaro}, we introduce time signature events \textbf{[TS]} to REMI. In this paper, we include five common time signatures, namely $\mathbf{\frac{2}{4}}$, $\mathbf{\frac{3}{4}}$, $\mathbf{\frac{4}{4}}$, $\mathbf{\frac{3}{8}}$, $\mathbf{\frac{6}{8}}$, in vocabulary. For other signatures, we refer to Table~\ref{ts_policy} to apply a simple conversion rule to preserve similar musical structure. 

\begin{table}[ht]
\renewcommand{\arraystretch}{1.5}
\caption{Original time signatures and their quantization strategies} 
\vspace{0.3cm}
\centering
\begin{tabular}{lr}
\toprule
Raw Time Signature & Policy\\
\midrule
$\mathbf{\frac{5}{4}}$      & convert to $\mathbf{\frac{2}{4}}$ bar + $\mathbf{\frac{3}{4}}$ bar  \\ \hline
$\mathbf{\frac{6}{4}}$      & split to two $\mathbf{\frac{3}{4}}$ bars  \\ \hline
$\mathbf{\frac{4}{8}}$      & quantize to $\mathbf{\frac{2}{4}}$  \\ \hline
$\mathbf{\frac{12}{8}}$     & split to two $\mathbf{\frac{6}{8}}$ bars  \\ \hline
all others & use $\mathbf{\frac{4}{4}}$ \\
\bottomrule
\end{tabular}
\label{ts_policy}
\end{table}

\subsubsection{Music Content}
We follow REMI~\citep{remi} to introduce several music content tokens in the input sequence. A \textbf{[B]} token denotes the start of each bar, followed by a bar-level time signature token \textbf{[TS]}. Beat tokens \textbf{[G]} are defined based on the time signature, ensuring that the resolution (12 grids for quarter-note (\musQuarter) and 6 for eighth-note (\musEighth)) can capture detailed rhythmic patterns. For example, the total number of beat tokens in the $\mathbf{\frac{4}{4}}$ is 48, and the number of beat tokens in $\mathbf{\frac{6}{8}}$ is 36.

Note tokens \textbf{[N]} are represented by two consecutive tokens: \textbf{[Note\_Pitch\_*]} (from A0 to C8) and \textbf{[Note\_Duration\_*]} (from \musThirtySecond \space to \musWhole), to present the pitch as well as the duration. 

\subsection{From Generality to Mastery}
Our training is conducted in two stages to balance general musical understanding with composer-specific nuance.

\subsubsection{Pre-training on General Musicality}
In the pre-training stage, we train our Transformer Decoder on a large, diverse corpus. The composer token is set to a uniform value (i.e., \textbf{[None]}), and the model learns to predict subsequent tokens based solely on the tempo $s$ condition and the prior context as:
\begin{align}
    p(\mathbf{x} \mid s, c=\text{None}) = \prod_{t=1}^{T} p(x_t \mid x_{<t}, s, c=\text{None}),
\end{align}
where $x_{<t}$ represents the tokens preceding time step $t$ after the composer token $c=\text{None}$ and the tempo token $s$. The goal here is to capture the general structure, motif development, and dynamics of music textures from a large corpus of music data without overfitting to any specific style.

\subsubsection{Fine-Tuning for Composer Mastery}
During fine-tuning, we update the global condition to include the actual composer token, and the probability function becomes:
\begin{equation}
    p(\mathbf{x} \mid s, c) = \prod_{t=1}^{T} p(x_t \mid x_{<t}, s, c)
\end{equation}
% Additionally, we introduce a lightweight style adapter module that injects composer-specific conditions into the hidden states of each decoder layer, similar to the typical bottleneck adapter used in NLP~\cite{adapter}. In particular, as shown in Figure [], we first concatenate the embedding vector of the composer with the hidden state, and then use a 2-layer MLP with GELU activation to let the model learn the style bias and down-project the input back to the shape of the hidden state. We follow typical practice to smooth the loss landscape, using a residual connection \cite{adapter}.  This module is designed to guide the model toward generating music that adheres to a particular composer’s style, which empirically slightly enhances the quality and expressiveness of the generated music. 

\begin{wrapfigure}{r}{0.45\textwidth}
    % \vspace{-2mm}
    \centering
    \includegraphics[width=\linewidth]{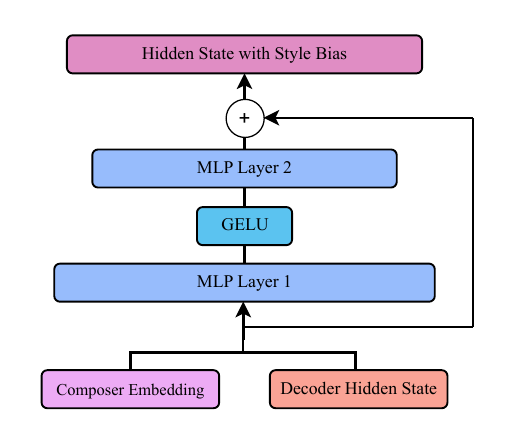}
    \caption{The architecture of the style adapter.}
    \label{fig:style_adapter}
    % \vspace{-5mm}
\end{wrapfigure}
Inspired by bottleneck adapters in NLP~\citep{adapter}, we insert a lightweight style adapter into each decoder layer to infuse composer‐specific information. As illustrated in Figure~\ref{fig:style_adapter}, we concatenate the composer embedding with the hidden state of the layer, feed the result through a two‐layer MLP (with a GELU nonlinearity) to learn a style bias, and project it back to the original hidden‐state dimension. A residual connection around this MLP stabilizes training~\citep{adapter}. Empirically, this adapter guides the model to generate music that more closely reflects the target composer's style, yielding modest improvements in quality and expressiveness.

The main difference between the two stages is that the pre-training ``generality" stage focuses on broad musical patterns, while the fine-tuning ``Mastery" stage focuses on composer identities to capture stylistic nuances.

% change location

\section{Experiments}
\subsection{Dataset and Preprocessing}
\subsubsection{Data Collection}
We collect a large corpus of symbolic music scores from nine existing datasets:  Pop909~\citep{pop909}, Pop1k7~\citep{cp}, Emopia~\citep{emopia}, DCML Corpora~\citep{dcmlab2020dcmlcorpora}, PIAST~\citep{bang-etal-2024-piast}, Kern Score~\citep{kernhumdrum}, classical piano midi datasets from \citep{drengskapur2021midi, pianomidi}, and a subset of PDMX~\citep{long2024pdmx}. We group them into three categories: pop, folk, and classical. After obtaining the classical collection, we separate the data of our four target composers, namely Bach, Mozart, Beethoven, and Chopin, from the dataset. We exclude these four composers in the ``Generality"  stage, and use them in the fine-tuning, and the rest in ``Mastery". Table \ref{data} presents the actual data quantity of each category or composer in both training stages.

Additionally, for pop music data, since there are some notes with overlong duration that spam 2-4 bars, we corrected these overlong notes to reduce the unexpected errors during the learning process. 

\begin{table}[t]
% \vspace{-3mm}
\centering
\caption{The summary of the training data in both stages. (a) presents the training data for the ``Generality" stage, and (b) presents the training data for the ``Mastery" stage.}
\begin{subtable}[b]{0.48\linewidth}
\renewcommand{\arraystretch}{1.0}
\centering
\caption{The summary of the pre-training datasets.}
    \begin{tabular}{lrr}
    \toprule
    Genre & \# Bars & \# Tokens \\
    \midrule
    Pop Music   & 298,740 &  36,820,539 \\
    Folk Songs  & 641,690 &  11,909,643 \\
    Classical   & 342,907 &  16,114,533 \\
    \midrule
    \textbf{Total} & 1,283,337 & 64,844,715\\
    \bottomrule
    \end{tabular}
    \label{data:pretrain_summary}
\end{subtable}
% \hfill
\begin{subtable}[b]{0.48\linewidth}
\centering
\caption{The summary of the fine-tuning datasets.}
    \begin{tabular}{lrr}
    \toprule
    Composer & \# unique pieces \\
    \midrule
    Bach &  307 \\
    Mozart &  161 \\
    Beethoven &  236 \\
    Chopin &  187 \\
    % Schumann &  80 \\
    \bottomrule
    \end{tabular}
    \label{data:unique_piece}
\end{subtable}

\label{data}
\end{table}

\subsubsection{Data Loading}
To balance the data quantity among the whole dataset, when sampling the data, we first sample the category with equal probability among the three categories (pop, folk, and classical) during the pre-training stage, and four composer categories with each composer as a category during the fine-tuning stage.

Then, we adopt a random segmentation by randomly selecting a segment of the input length as the input sequence from the complete song of the input data. To preserve music structure integrity, the segment must start and end at a certain bar in the original song, with some padding to make up the whole input length. For each segment, we prepend the composer and tempo tokens to make them global signals during the training process. And as mentioned in section 3.1.1, we include \textbf{[BOS]} and \textbf{[EOS]} only when it is the true start and end of a music. 

Finally, we conduct the data augmentation on note pitches. We augment the sequence by semitone transposition, ranging from [-3, 3], with upper and lower bounds of the key range.

\subsection{Model and Training}
\subsubsection{Model Architecture and Hyperparameters}
The model consists of 12 transformer decoder layers with a hidden size of 512 and the number of attention heads 8. The relative positional embedding~\cite{shaw2018self} is conducted through the training. We add five adapter modules after each even layer except the last layer, during the fine-tuning stage. The number of model parameters is around 46 million.

We pre-trained our model using a batch size of 8 and fine-tuned it using a batch size of 4, with a consistent context length of 2400. In both training stages, we used the Adam optimizer~\cite{kingma2014adam} with $\beta$ = (0.9, 0.999), while we conducted different learning schedulers. For the pre-training, we adopt a linear warm-up over 1000 steps to a maximum learning rate of 1e-4.  For the fine-tuning, we used a linear warm-up of 500 steps with a maximum learning rate of 1e-5. Both apply a cosine decay schedule over 500,000 steps, reducing the learning rate to $\frac{1}{10}$ of the respective maximum. For the main model modules, we clipped the gradient norms with a threshold of 0.5. For the adapter modules, we set this threshold to 2.0. The training pipeline is implemented in PyTorch on one NVIDIA RTX 3090 GPU, employing BF16 precision and gradient accumulation to simulate a batch size of 8 during pre-training. Empirically, the pre-training model reaches optimal performance after approximately 120,000 steps, and the fine-tuning model achieves its best performance after around 28,000 steps. We apply the nucleus
sampling~\citep{nucleus} when inference. We chose the sampling hyperparameters based on previous literature~\citep{emo, compembellish} as well as our experiments. We used $p$ = 0.99 for all models and use $\tau$ = 1.1 for composer-conditioned models (Mastery model and models for ablation) and $\tau$ = 1.0 for the generality model.

\subsection{Ablation Study and Baseline}
We conducted some ablation studies on the effectiveness of ``generality" pre-training stage before learning the specific traits of the composer style, namely: (1) We only used fine-tuning datasets to train the model from Scratch (as w/o. pre-training); and (2) We only used pre-training datasets to train the model without mapping to four composers (as w/o. fine-tuning). For fair comparison, we employ the same pitch augmentation strategy, keep all other hyperparameters the same as fine-tuning our Mastery model, and train to the best perceived quality, which takes around 52,000 steps for the ablation model (1).
We then compare its generated music quality and performance on style matching with the \textit{GnM} model. 

For the baselines, we chose NotaGen-large (0.5B parameters)~\cite{notagen} as a baseline. To ensure a fair and controlled comparison, we align the baseline with our two-stage training pipeline by using only the fine-tuning weights\footnote{\tiny weights\_notagen\_pretrain-finetune\_p\_size\_16\_p\_length\_1024\_p\_layers\_c\_layers\_6\_20\_h\_size\_1280\_lr\_1e-05\_batch\_1.pth} of NotaGen-large instead of the weights with reinforcement learning. Due to the lack of work on composer-style controllable music generation, there is no other baseline literature that includes the style of all four composers we are studying. Therefore, we chose both REMI~\citep{remi} and Emo-Disentanger~\citep{emo} as no-style conditioning baselines (and ablations).

\section{Objective Evaluation and Results}
We perform objective evaluations on both the composer style similarity and the music generation quality, on the four composers Bach, Mozart, Beethoven, and Chopin.

\subsection{Quantitative Music Quality}
To thoroughly evaluate the quality of generated music, we adopt a set of objective metrics that capture different musical dimensions from previous literature regarding symbolic music generation~\citep{compembellish}: pitch class entropy, grooving pattern similarity, and structureness indicators.

\subsubsection{Pitch Class Histogram Entropy}
For each segment, we calculate the 12-pitch class entropy:
\begin{equation}\label{eqn:pitch-entropy}
\mathcal{H}(\vectorize{h}) = - \sum_{i=0}^{11} h_i \log_2(h_i).
\end{equation}
A low entropy indicates a dominant tonality, whereas high entropy suggests ambiguity.

\subsubsection{Grooving Pattern Similarity}
We represent a bar's rhythm as a 64-dimensional binary vector $\vectorize{g}$ with 1 for a note onset. The similarity between two bars is given by:
\begin{equation}\label{eqn:groove-sim}
\mathcal{GS}(\vectorize{g}^a, \vectorize{g}^b) = 1 - \frac{1}{64}\sum_{i=0}^{63}\text{XOR}(g_i^a, g_i^b).
\end{equation}
Higher scores indicate consistent rhythmic patterns.

\subsubsection{Structureness Indicators}
Using the fitness scape plot $S \in [0,1]^{N \times N}$ from the self-similarity matrix, the structureness indicator over an interval $[l,u]$ is:
\begin{equation}\label{eqn:struct-indic}
\mathcal{SI}_l^u(S) = \max_{\substack{l \leq i \leq u\\ 1 \leq j \leq N}} S.
\end{equation}
This metric captures the strongest repetitive structure within a specified time scale, and we use 3, 6, and 9 seconds for short, mid, and long-term structureness.

\subsubsection{Quality Results}
We evaluate three objective metrics on generated samples from various models, including our proposed \textit{GnM} models, NotaGen, Emo-Disentangler, and the original REMI. From Table~\ref{fig:ob_quality}, we draw two key observations:
\begin{enumerate}[leftmargin=15pt,itemsep=1pt,topsep=2pt]
    \item Our proposed \textit{GnM} model achieves the highest pitch class entropy and nearly the best groove pattern similarity score. This demonstrates that the \textit{GnM} model, through its combination of "Generality" pre-training and "Mastery" fine-tuning, effectively learns both tonal and rhythmic patterns that closely align with real music data.
    \item We observe a decrease in the structureness indicator score from the \textit{GnM} ablation without pre-training to the final \textit{GnM} model. This implies that the version trained solely on limited composer-specific data tends to generate more repetitive structures, potentially due to overfitting on the smaller dataset. As a result, its generations may exhibit frequent repetition. In contrast, the final \textit{GnM} model, though showing a slightly lower structureness score, still performs comparably to other baselines. This may indicate that it captures more diverse mid-term and long-term structural variations.
\end{enumerate}

In summary, the proposed \textit{GnM} model demonstrates strong performance in capturing tonal and rhythmic qualities and maintains competitive structural coherence. These provide strong evidence of its generative quality and motivate further evaluations on style similarity and subjective tests.

\begin{table}[t]
\centering
\caption{The objective evaluation results among different music generation models and ablations.}
\vspace{3mm}
\resizebox{\textwidth}{!}{
\begin{tabular}{l|c|c|ccc}
\toprule
\multicolumn{1}{c|}{} & \multicolumn{1}{c|}{} &\multicolumn{1}{c|}{}& \multicolumn{3}{c}{SI} \\
Model & $\mathcal{H}$ & $\mathcal{GS}$ & $\mathcal{SI}_{\text{short}}$ & $\mathcal{SI}_{\text{mid}}$ & $\mathcal{SI}_{\text{long}}$ \\
\midrule
REMI~\citep{remi}                   & $\val{2.54}{0.31}$    & $\val{0.88}{0.05}$ & $\val{0.31}{0.05}$   & $\val{0.19}{0.08}$ & $\val{0.13}{0.10}$ \\
Emo~\citep{emo}                    & $\val{2.79}{0.28}$    & $\val{0.84}{0.04}$ & $\val{0.29}{0.03}$   & $\val{0.21}{0.06}$ & $\val{0.16}{0.07}$ \\
NotaGen~\citep{notagen}            & $\val{2.92}{0.26}$    & $\val{\textbf{0.97}}{\textbf{0.02}}$ & ${\val{0.35}{0.07}}$   & $\val{0.31}{0.10}$ & $\val{0.29}{0.13}$ \\
GnM (ours)     & $\val{\textbf{3.17}}{\textbf{0.18}}$    & $\val{0.96}{0.04}$ & $\val{0.31}{0.07}$   & $\val{0.23}{0.12}$ & $\val{0.20}{0.16}$ \\
GnM w/o. fine-tuning (ablation)              & $\val{2.86}{0.09}$    & $\val{0.93}{0.06}$ & $\val{0.30}{0.05}$   & $\val{0.27}{0.08}$ & $\val{0.22}{0.11}$ \\
GnM w/o. pre-training (ablation)                 & $\val{3.01}{0.18}$    & $\val{0.96}{0.04}$ & $\val{\textbf{0.41}}{\textbf{0.16}}$   & \textbf{$\val{\textbf{0.35}}{\textbf{0.17}}$} & \textbf{$\val{\textbf{0.33}}{\textbf{0.20}}$} \\

\bottomrule
\end{tabular}}
\label{fig:ob_quality}
\end{table}

\subsection{Quantitative Evaluation on Composer Styles}
\begin{figure}[t]
    \centering
    \includegraphics[width=\textwidth]{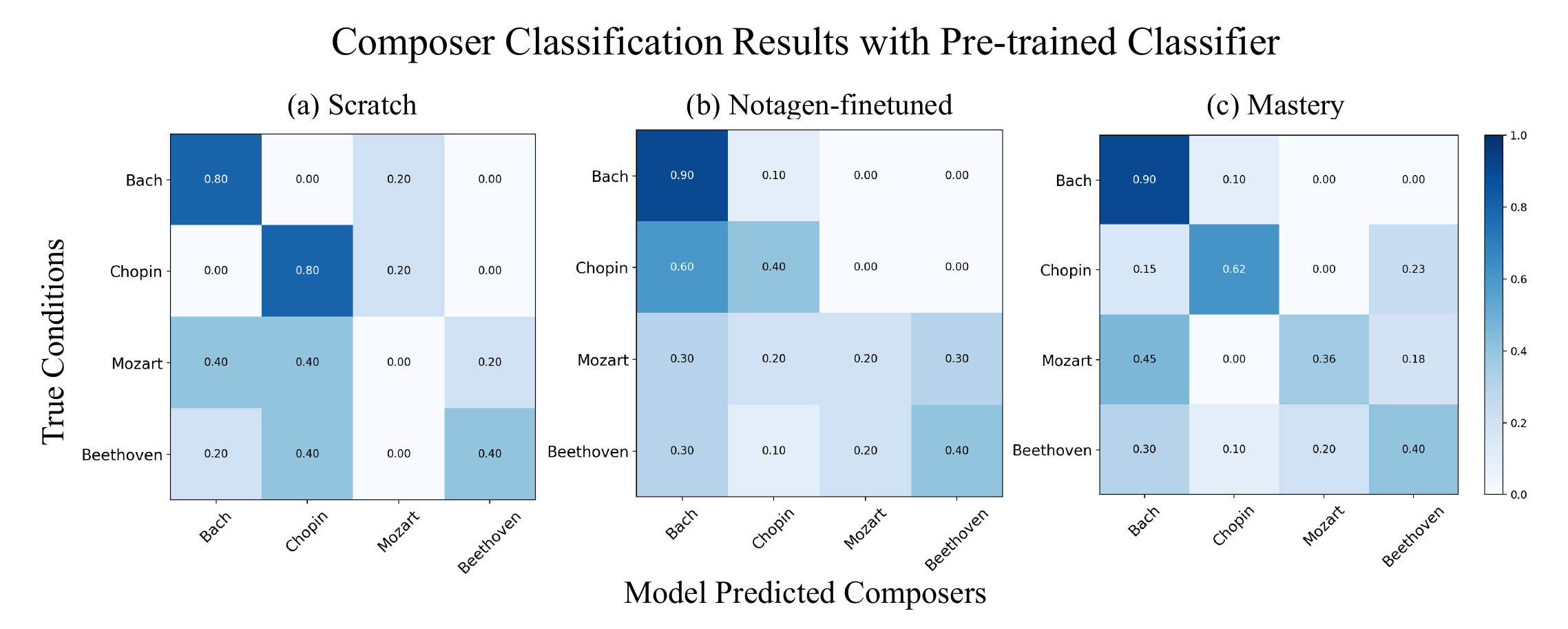}
    \caption{The composer classification accuracy on the \textit{GnM} model without the pre-training stage (as Scratch), fine-tuned NotaGen, and the final \textit{GnM} model (as Mastery). We exclude the Emo-Disentangler and the \textit{GnM} model without fine-tuning, because these two models cannot generate music with specific composer styles.}
    \label{fig:ob_acc}
\end{figure}

To evaluate the accuracy of the composer-style generation task, we conduct quantitative composer classification on samples generated by three models: the \textit{GnM} model without pre-training (referred as Scratch in Figure~\ref{fig:ob_acc}(a)), NotaGen, and the final \textit{GnM} model (referred as Mastery in Figure~\ref{fig:ob_acc}(c)). To ensure fairness, we train a composer classifier using a large classical music dataset~\citep{drengskapur2021midi}, focusing on the four target composers. We represent each piece using our extended REMI representation and use a similar transformer encoder architecture as the sequence model in recent SymRep~\citep{zhang2023symbolic}. The classifier achieves a validation accuracy of approximately 77.6\%. If this independently trained model can accurately classify the composer's style of generated samples, it indicates that the model has effectively captured the stylistic characteristics of the composers.

The classification results, shown in Figure~\ref{fig:ob_acc}, reveal that our proposed \textit{GnM} model (Mastery) outperforms the same-period state-of-the-art NotaGen in generating music in the style of Mozart and Beethoven. Interestingly, the Scratch model performs best for Chopin-style generation, but appears to underfit for Mozart-style music. These results suggest that the proposed \textit{GnM} model achieves more balanced and accurate composer-style generation, as evidenced by the classifier predictions.

\section{Subjective Evaluation and Results}
\begin{figure}[t]
    \centering
    \includegraphics[width=\textwidth]{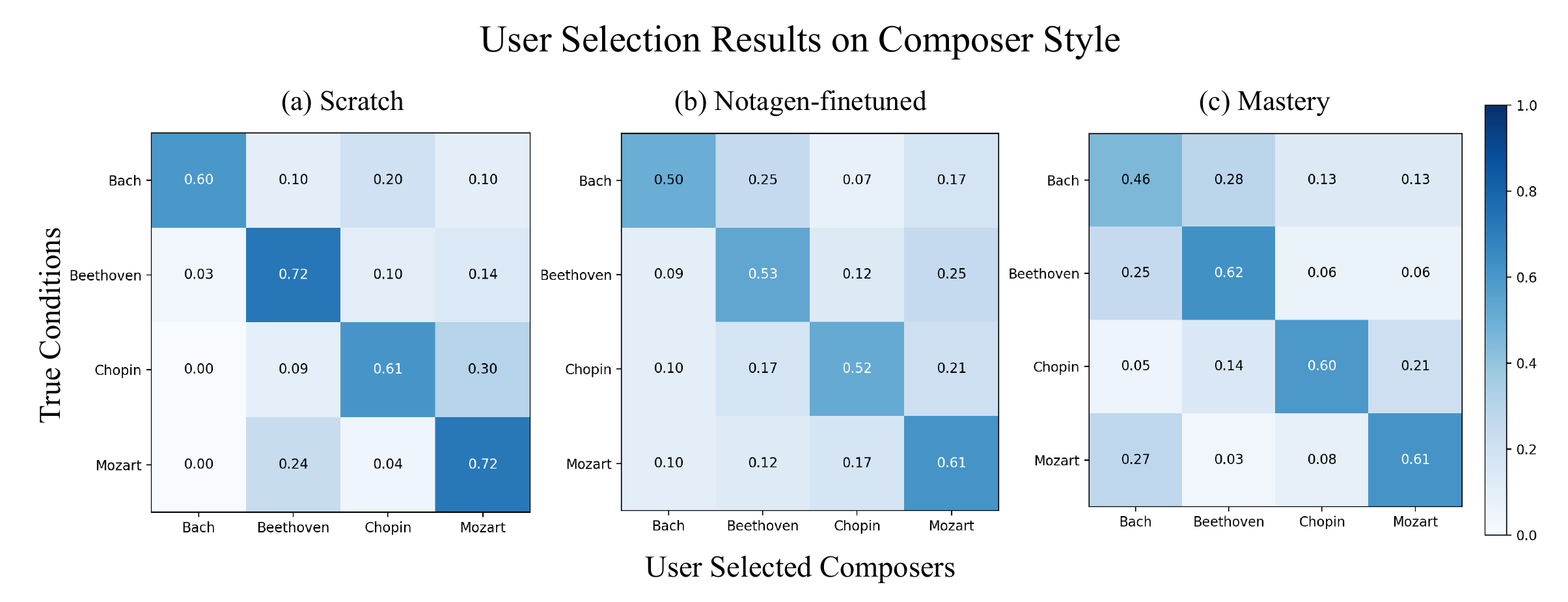}
    \caption{The composer style assessment from the subjective listening test, similar to the organization of quantitative classification results.}
    \label{fig:survey_acc}
    \vspace{-0.5cm}
\end{figure}

We conducted a subjective listening test via the online survey to assess the music quality as well as composer style similarities on the final \textit{GnM} model (Mastery), the \textit{GnM} model without pre-training (Scratch), the \textit{GnM} model without fine-tuning, NoteGen, and Emo-Disentangler. Considering the distinguishing styles of different classical composers can be difficult for people without professional music training, we employ a 2-choose-1 question format on composer selection. Particularly, we give an example audio for composers A and B, respectively, without stating their names. Then, we let participants listen to the generated sample, select the composer style that the generated sample matches the best, and rate the musicality of the generated sample from 1 to 5. In order to keep the model type blind to users, we randomly permute the order of the samples from five models for each question. Each participant will make 10 questions in total. In total, 54 people participated in the survey. We report our composer selection and musicality rating results in Figure~\ref{fig:survey_acc} and Table~\ref{tab:musicality_summary}, respectively.

% font to original, add reference on table
\begin{table}[t]
    \centering
    \caption{The musicality rating scores from the subjective listening test. We report the mean value, standard deviation (STD), standard error (SE), and 95\% Confidence Interval (CI).}
    \vspace{3mm}
    \begin{tabular}{l|cccc}
        \toprule
        Model    & Mean    & STD      & SE       & CI \\
        \midrule
        Emo-Disentanger~\citep{emo}      & 3.349 & 1.526  & 0.233 & 0.456 \\
        Notagen~\citep{notagen}  & 3.733 & 1.490  & 0.116 & 0.227 \\
        GnM (ours) (Mastery)  & 3.818 & 1.490  & \textbf{0.106} & \textbf{0.208} \\
        GnM w/o. fine-tuning & \textbf{3.824} & \textbf{1.290}  & 0.221 & 0.434 \\
        GnM w/o. pre-training (Scratch)   & 3.655 & 1.569  & 0.168 & 0.330 \\
        \bottomrule
    \end{tabular}
    \label{tab:musicality_summary}
    
\end{table}
From the results of the subjective listening test, both the \textit{GnM} Scratch and Mastery models produce samples with higher composer style similarities than Notagen on Beethoven, Chopin, and Mozart. The \textit{GnM} Scratch model achieves the best composer style similarities, but receives the lowest musicality rating from users. This further supports the conclusion in objective evaluations that it overfits on some composer styles but does not learn comprehensive musicality and presents high music quality. Notably, the Bach samples from Notagen and our Mastery model receive a trivial accuracy (50\% accuracy when randomly drawing between two composers), which reflects some challenges when balancing the composer style from the large-scale pre-training process, as well as the specific composer fine-tuning process.

\section{Discussion on Generality Pre-training and Mastery Fine-tuning}
To deeply understand the mechanisms by which models distinguish and emulate the styles of different composers, we performed comprehensive analyses focusing on inference behaviors, theoretical musical characteristics, and latent feature representations. Specifically, we compared our \textit{GnM} model (pre-trained only), \textit{GnM} Mastery model (fine-tuned after pre-training), and \textit{GnM} Scratch model (without pre-training) in various dimensions. Our inquiry was guided by three central questions:
\begin{itemize}[leftmargin=15pt,itemsep=1pt,topsep=2pt]
    \item \textbf{Q1:} Do the models differ in the musical note choices in the top-$p$ sampling inference process?
    \item \textbf{Q2:} How similar are the generations to the actual training data in chord progression structures?
    \item \textbf{Q3:} Does the Fréchet Audio Distance~\citep{fad} effectively capture differences in styles between the generations and training data?
\end{itemize}

We addressed these questions by generating 100 samples per composer using Mastery and Scratch models and 200 samples from the pre‑training‑only model, with 1000 samples in total, and employing the following analytical steps:
\begin{itemize}[leftmargin=15pt,itemsep=1pt,topsep=2pt]
    \item \textbf{Q1:} Given the first four bars of validation set pieces as primers, we quantified the diversity of model predictions for the subsequent bar. We used top-$p$ sampling with $p=0.99$, consistent with our inference setting, recording the remaining choices at each prediction step. We manage outliers by setting the minimum and maximum bounds at the 10th and 90th percentiles.
    \item \textbf{Q2:} We first extract chords using algorithm proposed in ~\citep{dai2020automatic}. Then, we use a sliding window (window size = 4; stride = 1) to identify chord progressions as tuples for all pieces of each composer and each model. Then, we measure the overlap ratios in the top 10, 15, and 20 common progressions between the Mastery/Scratch models and real data. We also employ the metrics commonly used in recommender systems, as mean Average Precision (mAP) and Normalized Discounted Cumulative Gain (NDCG)~\citep{ndcg}, to take progression frequency ranking into account. In order to remove bias introduced by pitch augmentation during training, \textbf{we shift the first chord in all progressions to have root pitch of C} and apply the same shift for the rest of the chords in each progression. We also treat progressions in patterns like A-B-A-B and B-A-B-A as the same progression to remove bias introduced by window size.
    \item \textbf{Q3:} We represented samples using pre-trained models (CLaMP 3~\citep{clamp3} and MIDI-Bert fine-tuned for composer classification~\citep{midibert}) to compute Fréchet Audio Distances between generated and real samples.
\end{itemize}

\subsection{Q1: Musical Note Choice}

\begin{wrapfigure}{r}{0.45\textwidth}
    % \vspace{-2mm}
    \centering
    \includegraphics[width=\linewidth]{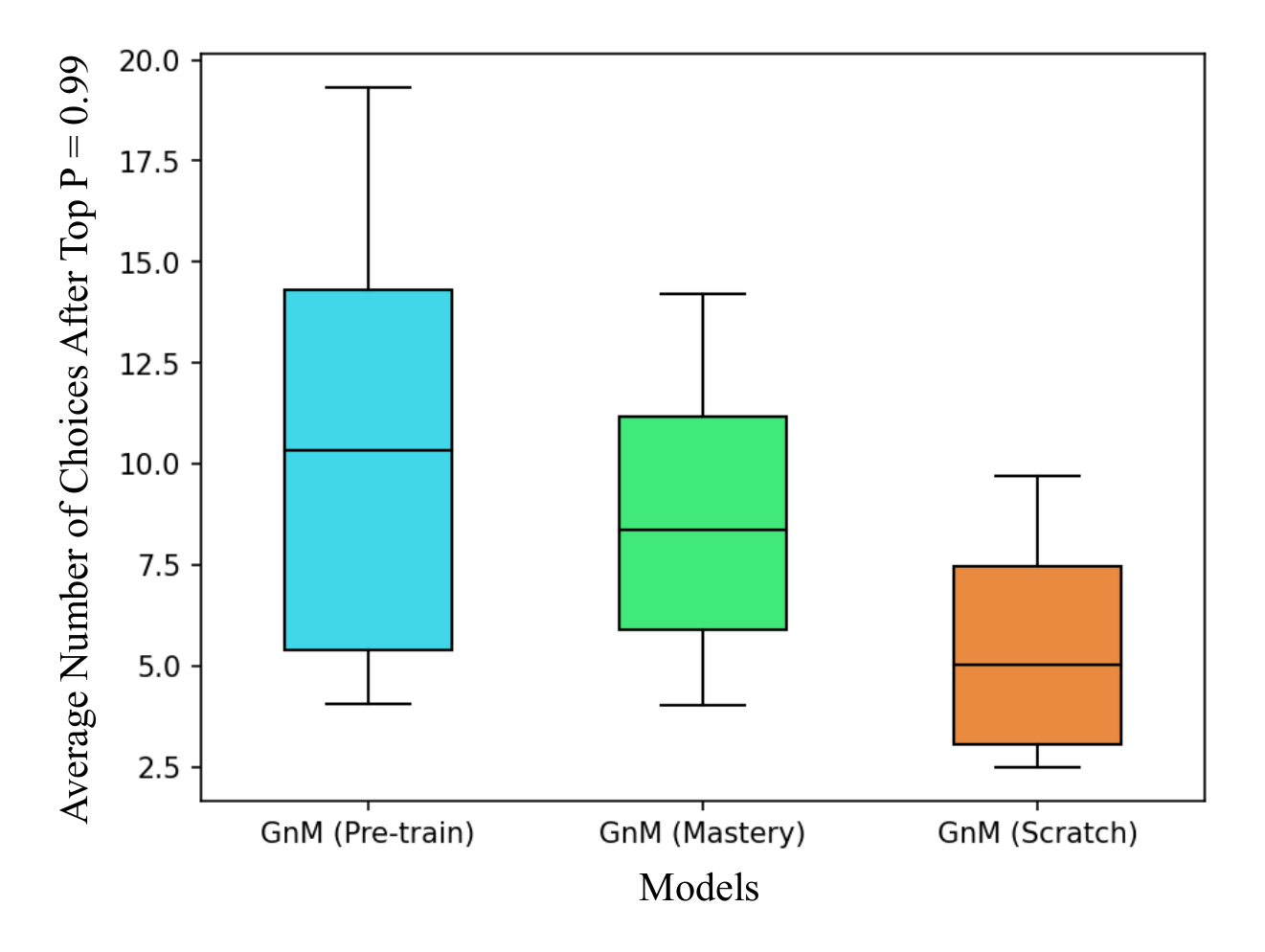}
    \caption{Average number of choices in each step when generating the 5th bar given the first 4 bars from validation sets.}
    \label{table:n_choice}
    % \vspace{-5mm}
\end{wrapfigure}

The results in Figure~\ref{table:n_choice} reveal a significant distinction in the number of generation choices across models. Both Mastery and Scratch display tighter inter-quartile ranges compared to the Generality model, indicating a more consistent selection process than the pre‑training‑only model. Notably, the Scratch model, without the ``generality" pre-training step, consistently shows fewer available note choices at all reported percentiles. This indicates limited musical knowledge and lower rhythmic complexity learned by the Scratch model. Conversely, the Mastery model, benefiting from extensive pre-training, offers more diverse and valid musical trajectories. The observed narrowing from generality to Mastery suggests composer-specific specialization, underscoring the role pre-training plays in equipping models with foundational musical knowledge that is subsequently refined during fine-tuning for specific style cues.

\subsection{Q2: Chord Progression Alignment}
Table~\ref{tab:comparison} presents the overlap percentages on Top-N chord progression between the Mastery/Scratch model and real data. It further demonstrates the distinction between Mastery and Scratch models in terms of the chord progression alignment. The Mastery model consistently presents significantly higher overlap percentages in top chord progressions when compared to real data across all composers, suggesting a superior capture of harmonic patterns in each composer's style. Moreover, from the perspective of information retrieval as ranking, the Mastery model outperforms the Scratch model on NDCG scores across all composer styles, and on mAP scores on Chopin and Mozart styles. This superior performance of the Mastery model can be attributed to its robust pre-trained musical foundation, which provides a rich foundation of general compositional patterns, then effectively guides the model away from overfitting to a composer imitation.

\begin{table}[t]
\centering
\resizebox{\textwidth}{!}{
\begin{tabular}{ll*{5}{C{1.5cm}}}
\toprule
\multirow{2}{*}{Composer} & \multirow{2}{*}{Model} &
  \multicolumn{3}{c}{Top‑N Progression Intersections} &
  \multicolumn{2}{c}{Metrics w/ Top‑20 Progressions}\\
\cmidrule(lr){3-5}\cmidrule(lr){6-7}
 & & Top 10 & Top 15 & Top 20 & mAP & NDCG \\
\midrule
\multirow{2}{*}{Bach}
 & Mastery  & \textbf{30.0\%} & \textbf{33.3\%} & \textbf{40.0\%} & 0.6386 & \textbf{0.4856} \\
 & Scratch  & 20.0\% & 26.7\% & 20.0\% & \textbf{0.7159} & 0.3219 \\
\midrule
\multirow{2}{*}{Chopin}
 & Mastery  & \textbf{70.0\%} & \textbf{46.7\%} & \textbf{60.0\%} & \textbf{0.9218} & \textbf{0.7085} \\
 & Scratch  & 40.0\% & 33.3\% & 25.0\% & 0.8767 & 0.3984 \\
\midrule
\multirow{2}{*}{Beethoven}
 & Mastery  & \textbf{40.0\%} & \textbf{33.3\%} & \textbf{50.0\%} & 0.7205 & \textbf{0.5852} \\
 & Scratch  & 40.0\% & 33.3\% & 35.0\% & \textbf{0.8588} & 0.4939 \\
\midrule
\multirow{2}{*}{Mozart}
 & Mastery  & \textbf{50.0\%} & \textbf{46.7\%} & \textbf{55.0\%} & \textbf{0.6579} & \textbf{0.5841} \\
 & Scratch  & 20.0\% & 20.0\% & 20.0\% & 0.5244 & 0.2827 \\
\bottomrule
\end{tabular}}
\vspace{0.3cm}
\caption{Comparison of chord‑progression statistics between real data and generated samples}
\label{tab:comparison}
\end{table}

On top of the overall statistic on chord progression patterns, we present another case study on the Top-10 chord progressions comparison on Bach and Chopin, to dive deeper into the progression difference between the Mastery and Scratch models. As presented in Table~\ref{tab:Bach_case_study} and Table~\ref{tab:Chopin_case_study}, each chord is annotated as \textbf{[Root:Type]} (e.g., F:M as F major chord). To reduce complexity, we simplify chord types into 7 main classes\footnote{\tiny major (M), minor (m), minor 7 (m7), dominant 7 (7), diminished (dim), augmented (aug), and suspended (sus).}. The detailed strategy of quality simplification can be referred to in our open-sourced implementation.

Table~\ref{tab:Bach_case_study} presents the Top 10 chord progressions in Bach style, including the real data, and samples from the Mastery and Scratch models. We use different colors to annotate shared chord progressions among the three columns. We observe that the Mastery model has achieved better progression alignment, in terms of ranking, than the Scratch model. Further, the chord quality distribution of the Mastery model is also closer to the real data, since the Scratch model has much more diverse chord qualities than the real data. For example, both real data and samples from the Mastery model mainly consist of major and minor chords, but augmented, sustained, and diminished chords appear frequently in the top frequent progressions, which strongly suggests the deprived learning of Bach pieces and potentially poor imitation or overfitting on shallow patterns.
% \begin{table}[t]
%   \centering

\begin{table}[t]
  \centering
  \caption{The top‑10 chord progressions on the Bach‑style real music data and model generations. Each color corresponds to one specific progression shared across multiple columns; all other progressions are shown in light gray.}
  \vspace{3mm}
  \begin{tabular}{ccc}
    \toprule
    \textbf{Real Data} & \textbf{Mastery}                       & \textbf{Scratch}                       \\
    \midrule
    \hlpurple{C:M - F:M - C:M - F:M}     & \hlpurple{C:M - F:M - C:M - F:M}       & \others{C:M - A:aug - C:M - A:aug}       \\
    \hlblue{C:M - G:M - C:M - G:M}     & \hlblue{C:M - G:M - C:M - G:M}       & \others{C:aug - D\#:M - C:aug - D\#:M}   \\
    \others{C:M - C:M - F:M - G:M}       & \hlgreen{C:M - D:m - C:M - F:M}       & \hlpurple{C:M - F:M - C:M - F:M}        \\
    \hlgreen{C:M - D:m - C:M - F:M}      & \others{C:M - C:M - D:M - G:M}        & \hlblue{C:M - G:M - C:M - G:M}        \\
    \others{C:m - F:M - A\#:M - D\#:M}   & \others{C:M - C:M - F:sus - B:m7}     & \others{C:M - G:sus - C:M - G:sus}       \\
    \others{C:M - F:M - C:sus - F:M}     & \others{C:M - D:M - C:M - G:M}        & \others{C:M - D:m - C:M - D:m}           \\
    \others{C:M - F:M - C:M - G:M}       & \others{C:m - A\#:M - D\#:M - F:M}    & \others{C:sus - F:M - C:sus - F:M}       \\
    \others{C:m - F:m - A\#:M - D\#:M}   & \others{C:M - D:m - C:M - D:m}        & \others{C:m - A\#:M - C:m - A\#:M}       \\
    \others{C:M - C:M - C:M - F:M}       & \others{C:M - D:m - C:M - G:M}        & \others{C:aug - F\#:M - C:m - C\#:M}     \\
    \others{C:M - F:M - A\#:M - F:M}     & \others{C:M - D:m - C:M - G:m}        & \others{C:dim - E:m - C:dim - E:m}       \\
    \bottomrule
  \end{tabular}
 
  \label{tab:Bach_case_study}
\end{table}

Another case study on Chopin samples and real data is illustrated in Table~\ref{tab:Chopin_case_study}. This case shows the better alignment more obviously, as the chord progression patterns of the Mastery model align exceptionally closely to the real data and still share a lot of traits in non-matched progressions. For example, the root pitches are generally F and G (except the normalized first root pitch), and the qualities are generally major or minor. However, the Scratch model has a worse matching rate with a larger difference in root pitch class. Samples from the Scratch model appear root pitch of D, A\# in a non-negligible frequency, which does not appear in the real data. It implies the underfitting of high-level features in Chopin pieces.

\begin{table}[t]
  \centering
  \caption{The top‑10 chord progressions on the Chopin‑style real music data and model generations.  Each color now corresponds to one specific progression shared across two or three columns; all other progressions are shown in light gray.}
\vspace{3mm}
  \begin{tabular}{ccc}
    \toprule
    \textbf{Real Data} & \textbf{Mastery}                     & \textbf{Scratch}                  \\
    \midrule
    \hlred{C:7 - F:M - C:7 - F:M}   & \hlgreen{C:7 - F:M - C:M - F:M}       & \hlpurple{C:M - F:M - C:M - F:M}      \\
    \hlblue{C:M - G:7 - C:M - G:7}   & \hlcyan{C:M - G:M - C:M - G:M}       & \others{C:m - A\#:m - C:m - A\#:m}      \\
    \hlgreen{C:7 - F:M - C:M - F:M}    & \hlpurple{C:M - F:M - C:M - F:M}       & \others{C:m - D:m - C:m - D:m}          \\
    \hlorange{C:M - G:7 - C:M - G:M}    & \hlred{C:7 - F:M - C:7 - F:M}       & \hlcyan{C:M - G:M - C:M - G:M}       \\
    \hlpurple{C:M - F:M - C:M - F:M}   & \hlorange{C:M - G:7 - C:M - G:M}       & \hlblue{C:M - G:7 - C:M - G:7}       \\
    \hlcyan{C:M - G:M - C:M - G:M}   & \hlblue{C:M - G:7 - C:M - G:7}       & \hlred{C:7 - F:M - C:7 - F:M}       \\
    \others{C:7 - F:m - C:7 - F:m}     & \others{C:M - C:M - F:M - G:M}         & \others{C:M - F:dim - C:M - G:M}        \\
    \others{C:m - F:m - C:m - G:7}     & \hlmagenta{C:M - F:M - C:M - G:7}       & \others{C:M - A\#:M - C:M - A\#:M}      \\
    \others{C:M - F:M - C:M - G:M}     & \others{C:m - G:7 - C:m - G:M}         & \others{C:M - D:M - C:M - D:M}          \\
    \hlmagenta{C:M - F:M - C:M - G:7}    & \others{C:M - G:M - D:7 - G:M}         & \others{C:M - C:M - F:M - G:M}          \\
    \bottomrule
  \end{tabular}
  \label{tab:Chopin_case_study}
\end{table}

\subsection{Q3: Fréchet Audio Distance}
The Fréchet Audio Distances computed using CLaMP 3 (Table~\ref{tab:clamp_fad}) and MIDI-Bert fine-tuned for composer classification (Table~\ref{tab:midibert_fad}) reinforce the insights derived from prior analysis. Generally, the Mastery model's samples yield lower distances to real training data across nearly all composers, suggesting more accurate stylistic replication. Notably, significant improvements are evident for Beethoven and Mozart when employing the MIDI-Bert embeddings, underscoring the value of leveraging models pre-trained for composer classification tasks. Interestingly, Bach exhibits slightly different patterns across embedding methods, suggesting potential variance in embedding sensitivity or inherent stylistic complexity. The higher variance observed with the from-Scratch model underlines its limited ability to capture high-level composer styles effectively.
\begin{table}[H]
    \centering
    \caption{Fréchet audio distance between training data and Mastery model and model from Scratch using CLaMP 3.}
  \vspace{3mm}
    \begin{tabular}{l|rrrr}
        \toprule
        Model    & Bach & Beethoven & Chopin &  Mozart       \\
        \midrule
        Mastery  & \textbf{442.98} & \textbf{547.30} & \textbf{523.63} & \textbf{468.64} \\
        {Scratch} & 451.01 & 598.22  & 549.02 & 558.26 \\
        \bottomrule
    \end{tabular}
    \label{tab:clamp_fad}
\end{table}

\begin{table}[H]
    \centering
    \caption{Fréchet audio distance between training data and Mastery model and model from Scratch using MIDI-Bert fine-tuned for composer classification.}
      \vspace{3mm}
    \begin{tabular}{l|rrrr}
        \toprule
        Model    & Bach & Beethoven & Chopin &  Mozart       \\
        \midrule
        Mastery  & 298.51 & \textbf{239.35} & \textbf{143.56} & \textbf{259.28} \\
        Scratch & \textbf{221.21} & 331.45  & 442.88 & 1548.26 \\
        \bottomrule
    \end{tabular}
    \label{tab:midibert_fad}
\end{table}

\subsection{Insights and Implications}
These findings collectively highlight the critical role of pre-training in symbolic music generation tasks. Pre-training provides models with a foundation of musical structures that enhances composer-specific fine-tuning by preventing overfitting and guaranteeing stylistic fidelity. In addition, the restricted choice space in composer-specific inference after fine-tuning suggests a compelling trade-off—choices are fewer, but they are more intensely associated with true compositional styles. Thus, our two-stage training approach improves not just the generative quality and diversity but also substantially captures composer-specific idiosyncrasies. Future work can investigate alternative methods of fine-tuning or investigate embedding-based metrics further for deeper style understanding.

\subsection{Limitations and Future Directions}
Despite promising outcomes, several limitations remain noteworthy. First, the composer-specific datasets' limited scale and inherent variability restrict the model's capacity to comprehensively generalize composer-specific attributes across varied compositional forms or musical forms. Secondly, using adapter modules for compositional conditioning, while beneficial for token-level stylistic control, might inadvertently constrain the long-term structural coherence due to excessive local stylistic bias. Lastly, even though current evaluation metrics capture objective stylistic aspects (e.g., chord progression), we leave subjective dimensions such as expressiveness, emotional nuances, and subtle stylistic distinctions insufficiently quantified.

Future research should prioritize dataset expansion, exploring alternative or hybrid conditioning approaches, and developing evaluation methodologies integrating professional musician judgments to enhance interpretative validity and subjective accuracy. Finally, answering the question of what elements collectively make a piece sound like a composer's style would be an important foundation for composer-specific music generation and information retrieval.

\section{Conclusion}
This research introduced a two-stage training scheme integrating large-scale general pre-training with composer-specific fine-tuning via adapter modules. Our in-depth analyses illustrate that pre-training greatly improves the capacity of models to mimic composer-specific styles and musicality over from-Scratch training or current baselines. Regardless of recognized limitations regarding dataset richness and evaluation depth, the Mastery model's higher performance highlights the significance of core and diverse musical knowledge in composer-conditioned symbolic music generation tasks. These results indicate exciting paths for future research, attempting to further increase compositional authenticity, increase dataset coverage, and better condition mechanisms, eventually approaching more advanced and human-oriented composer-style music generation.

\section{Ethical Considerations Regarding Musical Style Imitation}
While this research explores music generation by modeling the styles of classical composers whose works reside in the public domain, it is critical to address the broader ethical implications of such technological capabilities, especially concerning living artists. The ability to replicate and generate music in the unique stylistic signatures of musicians raises important questions around creative identity, intellectual property rights, and economic implications for contemporary creators.

Researchers and practitioners are encouraged to engage in ongoing dialogue, establishing clear ethical guidelines and legal frameworks to ensure that advancements in music generation technologies support rather than undermine artists' rights and contributions. Thus, transparent communication, responsible usage, and respect for the creative labor and identity of artists should be paramount considerations as this field continues to evolve.

\begin{ack}
We express special thanks to Jiaqing Zhang's participation in the discussion and data collection in the early stage of this work and Zhicheng Wang's help for data preprocessing in the early stage of research.

Thanks to my parents for their encouragement and Animenz's music for helping me pass through the hardest time for this work.
\end{ack}

\bibliographystyle{apalike}   
\bibliography{references}  

\begin{thebibliography}{}

\bibitem[Bang et~al., 2024]{bang-etal-2024-piast}
Bang, H., Choi, E., Finch, M., Doh, S., Lee, S., Lee, G.-H., and Nam, J. (2024).
\newblock {PIAST}: A multimodal piano dataset with audio, symbolic and text.
\newblock In Kruspe, A., Oramas, S., Epure, E.~V., Sordo, M., Weck, B., Doh, S., Won, M., Manco, I., and Meseguer-Brocal, G., editors, {\em Proceedings of the 3rd Workshop on NLP for Music and Audio (NLP4MusA)}, pages 5--10, Oakland, USA. Association for Computational Lingustics.

\bibitem[Burkholder et~al., 2010]{mushistory}
Burkholder, J.~P., Grout, D.~J., and Palisca, C.~V. (2010).
\newblock {\em A History of Western Music}.
\newblock W. W. Norton \& Company, New York, 8th edition.

\bibitem[Carter, 2002]{harmony}
Carter, E. (2002).
\newblock {\em Harmony Book}.
\newblock Carl Fischer Music.

\bibitem[Chen et~al., 2020]{chen2020musicsketchnet}
Chen, K., i~Wang, C., Berg-Kirkpatrick, T., and Dubnov, S. (2020).
\newblock Music sketchnet: Controllable music generation via factorized representations of pitch and rhythm.
\newblock In {\em Proceedings of the 21th International Society for Music Information Retrieval Conference, {ISMIR}}.

\bibitem[Chou et~al., 2021]{midibert}
Chou, Y.-H., Chen, I.-C., Chang, C.-J., Ching, J., and Yang, Y.-H. (2021).
\newblock {MidiBERT-Piano}: Large-scale pre-training for symbolic music understanding.
\newblock {\em arXiv preprint arXiv:2107.05223}.

\bibitem[Dai et~al., 2020]{dai2020automatic}
Dai, S., Zhang, H., and Dannenberg, R.~B. (2020).
\newblock Automatic analysis and influence of hierarchical structure on melody, rhythm and harmony in popular music.
\newblock In {\em Proc. AIMC}.

\bibitem[{DCMLab}, 2023]{dcmlab2020dcmlcorpora}
{DCMLab} (2023).
\newblock dcml\_corpora.
\newblock \url{https://github.com/DCMLab/dcml_corpora}.
\newblock GitHub repository, accessed March 27, 2025.

\bibitem[Drengskapur, 2024]{drengskapur2021midi}
Drengskapur (2024).
\newblock Midi classical music dataset.
\newblock \url{https://huggingface.co/datasets/drengskapur/midi-classical-music}.
\newblock Hugging Face dataset; accessed March 27, 2025.

\bibitem[Hadjeres et~al., 2017]{deep_bach}
Hadjeres, G., Pachet, F., and Nielsen, F. (2017).
\newblock {D}eep{B}ach: a steerable model for {B}ach chorales generation.
\newblock In {\em Proc. ICLR}, pages 1362--1371.

\bibitem[Holtzman et~al., 2020]{nucleus}
Holtzman, A., Buys, J., Du, L., Forbes, M., and Choi, Y. (2020).
\newblock The curious case of neural text degeneration.
\newblock In {\em Proc. ICLR}.

\bibitem[Houlsby et~al., 2019]{adapter}
Houlsby, N., Giurgiu, A., Jastrzebski, S., Morrone, B., De~Laroussilhe, Q., Gesmundo, A., Attariyan, M., and Gelly, S. (2019).
\newblock Parameter-efficient transfer learning for {NLP}.
\newblock In {\em Proc. ICML}.

\bibitem[Hsiao et~al., 2021]{cp}
Hsiao, W., Liu, J., Yeh, Y., and Yang, Y.-H. (2021).
\newblock {Compound Word Transformer}: Learning to compose full-song music over dynamic directed hypergraphs.
\newblock In {\em Proc. AAAI}.

\bibitem[Huang et~al., 2019]{musictransformer}
Huang, C.-Z.~A., Vaswani, A., Uszkoreit, J., Simon, I., Hawthorne, C., Shazeer, N., Dai, A.~M., Hoffman, M.~D., Dinculescu, M., and Eck, D. (2019).
\newblock {Music Transformer}: Generating music with long-term structure.
\newblock In {\em Proc. ICLR}.

\bibitem[Huang et~al., 2024]{emo}
Huang, J., Chen, K., and Yang, Y.-H. (2024).
\newblock Emotion-driven piano music generation via two-stage disentanglement and functional representation.
\newblock In {\em Proc. ISMIR}.

\bibitem[Huang and Yang, 2020]{remi}
Huang, Y.-S. and Yang, Y.-H. (2020).
\newblock {Pop Music Transformer}: Beat-based modeling and generation of expressive pop piano compositions.
\newblock In {\em Proc. ACM Multimed.}

\bibitem[Hung et~al., 2021]{emopia}
Hung, H., Ching, J., Doh, S., Kim, N., Nam, J., and Yang, Y.-H. (2021).
\newblock {EMOPIA:} {A} multi-modal pop piano dataset for emotion recognition and emotion-based music generation.
\newblock In {\em Proc. ISMIR}.

\bibitem[Järvelin and Kekäläinen, 2002]{ndcg}
Järvelin, K. and Kekäläinen, J. (2002).
\newblock Cumulated gain-based evaluation of ir techniques.
\newblock {\em ACM Trans. Inf. Syst.}, 20:422--446.

\bibitem[Kilgour et~al., 2019]{fad}
Kilgour, K., Zuluaga, M., Roblek, D., and Sharifi, M. (2019).
\newblock Fréchet audio distance: A metric for evaluating music enhancement algorithms.
\newblock {\em Proc. INTERSPEECH}.

\bibitem[Kingma and Ba, 2015]{kingma2014adam}
Kingma, D.~P. and Ba, J. (2015).
\newblock Adam: A method for stochastic optimization.
\newblock In {\em Proc. ICLR}.

\bibitem[Krueger, 1996]{pianomidi}
Krueger (1996).
\newblock Classical piano music page.
\newblock \url{http://piano-midi.de/}.
\newblock Accessed March 27, 2025.

\bibitem[Long et~al., 2025]{long2024pdmx}
Long, P., Novack, Z., Berg-Kirkpatrick, T., and McAuley, J. (2025).
\newblock Pdmx: A large-scale public domain musicxml dataset for symbolic music processing.
\newblock In {\em ICASSP 2025 - 2025 IEEE International Conference on Acoustics, Speech and Signal Processing (ICASSP)}, pages 1--5.

\bibitem[Lu et~al., 2023]{musecoco}
Lu, P., Xu, X., Kang, C., Yu, B., Xing, C., Tan, X., and Bian, J. (2023).
\newblock {MuseCoco}: Generating symbolic music from text.
\newblock {\em CoRR}, abs/2306.00110.

\bibitem[{OpenAI}, 2019]{openai2019musenet}
{OpenAI} (2019).
\newblock Musenet.
\newblock Accessed: March 27, 2025.

\bibitem[Pfleiderer et~al., 2017]{Pfleiderer:2017:BOOK}
Pfleiderer, M., Frieler, K., Abe{\ss}er, J., Zaddach, W.-G., and Burkhart, B., editors (2017).
\newblock {\em {I}nside the {J}azzomat - {N}ew {P}erspectives for {J}azz {R}esearch}.
\newblock Schott Campus.

\bibitem[Roberts et~al., 2018]{roberts2018musicvae}
Roberts, A., Engel, J., Raffel, C., Simon, I., and Hawthorne, C. (2018).
\newblock Musicvae: Creating a palette for musical scores with machine learning.
\newblock \url{https://magenta.tensorflow.org/music-vae}.
\newblock Accessed: 2025-03-29.

\bibitem[Sapp, 2005]{kernhumdrum}
Sapp, C.~S. (2005).
\newblock Online database of scores in the humdrum file format.
\newblock In {\em Proc. ISMIR}.

\bibitem[Shaw et~al., 2018]{shaw2018self}
Shaw, P., Uszkoreit, J., and Vaswani, A. (2018).
\newblock Self-attention with relative position representations.
\newblock In {\em Proc. North American Chapter of the Association for Computational Linguistics: Human Language Technologies}.

\bibitem[Vaswani et~al., 2017]{transformer}
Vaswani, A., Shazeer, N., Parmar, N., Uszkoreit, J., Jones, L., Gomez, A.~N., Kaiser, L., and Polosukhin, I. (2017).
\newblock Attention is all you need.
\newblock In {\em Proc. NeurIPS}.

\bibitem[von Rütte et~al., 2023]{figaro}
von Rütte, D., Biggio, L., Kilcher, Y., and Hofmann, T. (2023).
\newblock Figaro: Controllable music generation using learned and expert features.
\newblock In {\em Proc. ICLR}.

\bibitem[Wang et~al., 2025]{notagen}
Wang, Y., Wu, S., Hu, J., Du, X., Peng, Y., Huang, Y., Fan, S., Li, X., Yu, F., and Sun, M. (2025).
\newblock Notagen: Advancing musicality in symbolic music generation with large language model training paradigms.

\bibitem[Wang et~al., 2020a]{pop909}
Wang, Z., Chen, K., Jiang, J., Zhang, Y., Xu, M., Dai, S., and Xia, G. (2020a).
\newblock {POP909:} {A} pop-song dataset for music arrangement generation.
\newblock In {\em Proc. ISMIR}.

\bibitem[Wang and Xia, 2021]{wang2021musebert}
Wang, Z. and Xia, G. (2021).
\newblock Musebert: Pre-training of music representation for music understanding and controllable generation.
\newblock In {\em Proc. ISMIR}.

\bibitem[Wang et~al., 2020b]{wang2020pianotree}
Wang, Z., Zhang, Y., Zhang, Y., Jiang, J., Yang, R., Xia, G., and Zhao, J. (2020b).
\newblock Pianotree vae: Structured representation learning for polyphonic music.
\newblock In {\em Proc. ISMIR}, pages 368--375.

\bibitem[Wu et~al., 2025]{clamp3}
Wu, S., Guo, Z., Yuan, R., Jiang, J., Doh, S., Xia, G., Nam, J., Li, X., Yu, F., and Sun, M. (2025).
\newblock Clamp 3: Universal music information retrieval across unaligned modalities and unseen languages.

\bibitem[Wu and Yang, 2023]{compembellish}
Wu, S.-L. and Yang, Y.-H. (2023).
\newblock {Compose \& Embellish}: Well-structured piano performance generation via a two-stage approach.
\newblock In {\em Proc. ICASSP}.

\bibitem[Yang et~al., 2019]{yang2019deep}
Yang, R., Wang, D., Wang, Z., Chen, T., Jiang, J., and Xia, G. (2019).
\newblock Deep music analogy via latent representation disentanglement.
\newblock In {\em Proc. ISMIR}, pages 596--603.

\bibitem[Zeng et~al., 2021]{musicbert}
Zeng, M., Tan, X., Wang, R., Ju, Z., Qin, T., and Liu, T.-Y. (2021).
\newblock Musicbert: Symbolic music understanding with large-scale pre-training.
\newblock In {\em Proc. ACL}.

\bibitem[Zhang et~al., 2023]{zhang2023symbolic}
Zhang, H., Karystinaios, E., Dixon, S., Widmer, G., and Cancino-Chac{\'o}n, C.~E. (2023).
\newblock Symbolic music representations for classification tasks: A systematic evaluation.
\newblock In {\em Proc. ISMIR}.

\end{thebibliography}

\end{document}